\documentclass[reprint,prb,showpacs]{revtex4-1}
\usepackage{amsmath,amsthm,amssymb,amscd,float,accents,enumerate}
\usepackage{mathrsfs}
\usepackage{graphicx}
\usepackage[T1]{fontenc}
\usepackage{lmodern}
\usepackage{hyperref}
\usepackage{color}
\newcommand{\al}{\alpha}
\newcommand{\be}{\beta}
\newcommand{\de}{\delta}

\newcommand{\vep}{\varepsilon}
\newcommand{\ga}{\gamma}
\newcommand{\ka}{\kappa}

\renewcommand{\th}{\theta}

\newcommand{\De}{\Delta}

\newcommand{\La}{\Lambda}

\newcommand{\bde}{\boldsymbol{\delta}}

\newcommand{\bk}{\mathbf{k}}

\newcommand{\bn}{\mathbf{n}}

\newcommand{\bs}{\mathbf{s}}
\newcommand{\bt}{\mathbf{t}}
\newcommand{\bT}{\mathbf{T}}

\newcommand{\bx}{\mathbf{x}}

\newcommand{\bS}{\mathbf S}
\newcommand{\bmu}{{\boldsymbol{\mu}}}

\newcommand{\bphi}{{\boldsymbol{\phi}}}

\newcommand{\bxi}{{\boldsymbol{\xi}}}

\newcommand{\tK}{\widetilde{K}}

\newcommand{\CC}{{\mathbb C}}

\newcommand{\ZZ}{{\mathbb Z}}
\newcommand{\cB}{{\mathcal B}}

\newcommand{\cE}{{\mathcal E}}

\newcommand{\cH}{{\mathcal H}}

\newcommand{\cP}{{\mathcal P}}

\newcommand{\cZ}{{\mathcal Z}}
\newcommand{\bj}{\bar\jmath}
\newcommand{\pd}{\partial}

\newcommand{\id}{1\hspace{-.25em}{\rm I}}

\newcommand{\ket}[1]{|#1\rangle}

\newcommand{\ms}{\mspace{1mu}}

\renewcommand{\le}{\leqslant}
\renewcommand{\ge}{\geqslant}
\newcommand{\tends}[1]{\bbuildrel{\hbox to 2em{\rightarrowfill}}_{#1}^{}}
\newcommand{\tr}{\operatorname{tr}}

\renewcommand{\Re}{\operatorname{Re}}
\newcommand{\iu}{\mathrm i}
\newcommand{\e}{\mathrm e}
\newcommand{\Or}{\mathrm{O}}

\newcommand{\su}{\mathrm{su}}

\newcommand{\en}{\enspace}

\newcommand{\pdf}[2]{\frac{\partial #1}{\partial #2}}

\begin{document}
\title{Integrable open spin chains related to infinite matrix product states}
\author{B. \surname{Basu-Mallick}}\email{bireswar.basumallick@saha.ac.in} \affiliation{Theory
  Division, Saha Institute of Nuclear Physics, 1/AF Bidhan Nagar, Kolkata 700 064, India}
\author{F. \surname{Finkel}}\email{ffinkel@ucm.es} \author{A. \surname{Gonz\'alez-L\'opez}}%
\email[Corresponding author. Email address: ]{artemio@ucm.es} \affiliation{Departamento de F\'\i sica
  Te\'orica II, Universidad Complutense de Madrid, 28040 Madrid, Spain}
\date{April 8, 2016}
\begin{abstract}
  In this paper  we study an~$\su(m)$-invariant open version of the Haldane--Shastry spin chain
  whose ground state can be obtained from the chiral correlator of the~$c=m-1$ free boson boundary
  conformal field theory. We show that this model is integrable for a suitable choice of the chain
  sites depending on the roots of the Jacobi polynomial~$P_N^{\be-1,\be'-1}$, where $N$ is the
  number of sites and $\be,\be'$ are two positive parameters. We also compute in closed form the
  first few nontrivial conserved charges arising from the twisted Yangian invariance of the model.
  We evaluate the chain's partition function, determine the ground state energy and deduce a
  complete description of the spectrum in terms of Haldane's motifs and a related classical vertex
  model. In particular, this description entails that the chain's level density is normally
  distributed in the thermodynamic limit. We also analyze the spectrum's degeneracy, proving that
  it is much higher than for a typical Yangian-invariant model.
\end{abstract}
\pacs{75.10.Pq, 02.30.Ik, 11.25.Hf}
\maketitle
\section{Introduction}
Recent experiments involving optical lattices of ultracold Rydberg atoms and trapped ions, neutral
atoms in optical cavities, etc., offer the possibility of realizing various theoretical models of
lower-dimensional spin systems with long-range interactions in a remarkably precise way
\cite{SZFHC15,RGLSS14,JLHHZ14,KCIKD09,PC04}. For example, by using hyperfine `clock' states of
trapped ${}^{171}\mathrm{Yb}^+$ ions it has become possible to realize one-dimensional spin
systems with tunable long-range interactions, where the coupling between the $i$-th and $j$-th
lattice sites falls off approximately algebraically as $J_{ij} \propto 1/|i-j|^\alpha$, with
$\alpha \in (0,3)$. Furthermore, it has been found that, unlike the case of spin chains with
nearest or next-to-nearest neighbor interactions, spin chains with long-range interactions often
exhibit interesting physical phenomena like realization of quantum spin glasses, quantum crystals
and high-speed propagation of correlations exceeding light-cone-like
bounds~\cite{LZ13,WLPB08,RGLSS14,JLHHZ14}.

Due to these salient features of strongly correlated lower-dimensional systems, the theoretical
investigation of exactly solvable and quantum integrable spin chains with long-range interactions
has acquired great impetus. The study of this type of quantum integrable spin systems was
pioneered by Haldane and Shastry~\cite{Ha88,Sh88}, who found the exact spectrum of a circular
array of equispaced $\su(2)$ spins with two-body interactions inversely proportional to the square
of their chord distances. This Haldane--Shastry (HS) spin chain has many remarkable properties: to
name only a few, its exact ground state wave function coincides with the $U\rightarrow \infty $
limit of Gutzwiller's variational wave function for the Hubbard model~\cite{GV87,GJR87}, and its
spinon excitations obey a generalized Pauli exclusion principle~\cite{Ha91b}. Furthermore, this
chain exhibits Yangian quantum group symmetry, due to which the corresponding spectrum can be
expressed in closed form by using the so called `motifs' \cite{HHTBP92,BGHP93}.

In the past few years, infinite matrix product states (MPS) related to (1+1)-dimensional conformal
field theories (CFT) have been used to construct HS-like quantum spin chains with $\su(m)$ spin
and periodic boundary conditions~\cite{CS10,NCS11,NCS12,BQ14}. In this approach,
finite-dimensional matrices associated with MPS are replaced by chiral vertex operators of a CFT,
and the corresponding correlator is interpreted as the ground state wave function of the spin
system. Very recently, an inhomogeneous open version of the HS spin chain has been constructed by
using infinite MPS from a suitable boundary conformal field theory~\cite{TS15}. In the $\su(2)$
case this construction naturally yields a linear system determining the two-point spin correlation
functions, which can be solved in closed form for a particular (uniformly distributed) choice of
the chain sites. In fact, the previous reference mainly focus on three instances of chains with
equispaced sites, for which they discuss the integrability, conjecture a formula for the spectrum
from numerical computations and determine the twisted Yangian generators responsible for the huge
spectral degeneracy. The purpose of the present paper is twofold. In the first place, we shall
show that these three equispaced chains can be embedded into a large class of integrable open spin
chains whose lattice sites (no longer equally spaced) depend on two free parameters. We shall then
compute the first few nontrivial conserved charges arising from the model's twisted Yangian
invariance, and show that they coincide with the twisted Yangian generators of
Ref.~\onlinecite{TS15} in the three equispaced cases. Secondly, we shall evaluate in closed form
the partition function of these models, providing a rigorous derivation of the formula for the
energy spectrum conjectured in Ref.~\onlinecite{TS15} for the equispaced cases. It should be
stressed that our results apply to the whole two-parameter family of integrable spin chains
mentioned above, and not just to the three particular instances thereof studied
in~Ref.~\onlinecite{TS15}. In particular, in the~$\su(2)$ case the two-point spin correlators of all
of these models are determined by the linear system deduced in the latter reference for the three
equispaced cases.

The paper is organized as follows. Section~\ref{sec.SA} is devoted to recalling the definition of
the $\su(m)$ Simons--Altshuler model and its connection with the $c=m-1$ free boson boundary CFT.
In Section~\ref{sec.IG} we construct a two-parameter integrable generalization of the latter
model, and show that its ground state is still given by a chiral correlator of a boundary CFT with
$m-1$ free bosons. The derivation of the first few nontrivial conserved charges arising from the
twisted Yangian symmetry of the model is presented in Section~\ref{sec.tYs}. In
Section~\ref{sec.PF} we show how to compute in closed form the model's partition function, from
which we deduce its equivalence to a one-dimensional classical vertex model with a simple
dispersion relation. This result is used in Section~\ref{sec.GSS} to provide a description of the
spectrum in terms of Haldane's motifs. In particular, we compute the ground state energy and prove
the formula for the energy spectrum proposed in Ref.~\onlinecite{TS15} for the three equispaced
cases. In Section~\ref{sec.CO} we present a brief summary of our work and outline several possible
future developments thereof. The paper ends with a short technical appendix, in which we prove an
identity for the zeros of Jacobi polynomials needed to compute the twisted Yangian conserved
charges.

\section{The \boldmath$\su(m)$ Simons--Altshuler model}\label{sec.SA}
Consider, to begin with, a spin $1/2$ chain whose fixed sites $z_j=\e^{2\iu\xi_j}$
($\xi_j\in[0,\pi/2]$, $j=1,\dots,N$) lie on the upper unit circle in the complex plane, and let
$\ket{s_j}$ ($s_j=\pm1$) be the canonical spin basis at the $j$-th site. We shall denote the mirror
image~$z_j^*$ of the lattice site~$z_j$ by $z_{\bj}$, and shall also set
\[
u_j=\frac12\,(z_j+z_{\bj})=\cos(2\xi_j)\,.
\]
Following Ref.~\onlinecite{TS15}, we shall take as ground state of the model under consideration
the chiral correlator
\begin{equation}\label{psi}
\psi(s_1,\dots,s_N)=\langle A^{s_1}(u_1)A^{s_2}(u_2)\cdots A^{s_N}(u_N)\rangle\,,
\end{equation}
where $A^{s_j}(u_j)=\chi_j:\e^{\iu s_j \phi(u_j)/\sqrt 2}:$ ($:\cdots:$ denoting, as usual, the
normal ordering), $\phi(u)$ is a chiral bosonic field from the $c=1$ free boson CFT, and
$\chi_j=s_j$ for even~$j$ and $1$ otherwise. As shown in Ref.~\onlinecite{TS15}, $\psi$ is
annihilated by the $3N$ operators~\footnote{Here and in what follows sums and products over the
  indices $i,j,k,l$ range from~$1$ to~$N$ unless otherwise stated, and $\sum_{i\ne j}$,
  $\sum_{i(\ne j)}$ respectively denote summation over the two indices~$i,j$ and over the single
  index~$i$ with the restriction $i\ne j$.}
\[
\La^a_k=\frac1{3\sqrt2}\sum_{j(\ne
  k)}(w_{kj}+w_{k\bar\jmath})\Big(s^a_j+\iu\sum_{b,c=1}^3\vep_{abc}s_k^bs_j^c\Big)\,,
\]
with $1\le k\le N$ and $a=1,2,3$, where $\bs_i=(s_i^1,s_i^2,s_i^3)$ is the spin operator of the
$i$-th particle, $\vep_{abc}$ is the Levi-Civita symbol,
and we have set
\[
w_{kj}=\frac{z_k+z_j}{z_k-z_j}\,.
\]
Thus the state~$\psi(s_1,\dots,s_N)$ defined above is by construction the ground state of the
positive-definite Hamiltonian~$\sum_{a,i}\big(\La^a_i\big)^\dagger\La^a_i$. Calling
\[
h_{ij}=w_{ij}(c_i-c_j)+w_{i\bj}(c_i+c_j)\,,\quad c_j=w_{\bj j}+\sum_{l(\ne j)}(w_{lj}+w_{\bar l j})\,,
\]
it can be proved~\cite{TS15} that the latter Hamiltonian can be written as
\begin{equation}\label{Hgen}
  \cH_{\mathrm{MPS}} = \sum_{i\ne j}\bigg(\frac1{|z_i-z_j|^2}+\frac1{|z_i-z_{\bj}|^2}-\frac{h_{ij}}{12}\bigg)\bs_i\cdot \bs_j
\end{equation}
up to an additive constant and an unimportant term proportional to the square of the total spin
operator~$\bS\equiv\sum_i\bs_i$. For the three particular cases in which the angles of the chain
sites are given by
\begin{equation}\label{unifc}
\xi_j=\frac{\pi(j-\frac12)}{2N},\en\frac{\pi j}{2N+2},\en\frac{\pi j}{2N+1}\,,
\end{equation}
it is shown in Ref.~\onlinecite{TS15} that the term~$h_{ij}$ in Eq.~\eqref{Hgen} is a constant
(independent of $i$ and $j$) respectively equal to $0$, $4$, $2$. In these so called
\emph{uniform} cases, the model~\eqref{Hgen} essentially coincides with the integrable open chain
of Haldane--Shastry type introduced by Simons and Altshuler~\cite{SA94}, whose
Hamiltonian is usually written as~\cite{BPS95}
\begin{align}\label{HBC}
  \cH&= \sum_{i\ne j}\bigg(\frac1{|z_i-z_j|^2}+\frac1{|z_i-z_{\bj}|^2}\bigg)(P_{ij}-1)\notag\\
  &=\frac14\sum_{i\ne j}\Big(\sin^{-2}\xi_{ij}^-
  +\sin^{-2}\xi_{ij}^+\Big)(P_{ij}-1)\,.
\end{align}
In the previous formula $\xi_{ij}^\pm=\xi_i\pm\xi_j$, with~$\xi_j$ given by Eq.~\eqref{unifc}, and
$P_{ij}$ is the spin exchange operator, defined by
\[
P_{ij}\ket{\dots,s_i,\dots,s_j,\dots}=\ket{\dots,s_j,\dots,s_i,\dots}\,.
\]
Indeed, since $P_{ij}=2\,\bs_i\cdot\bs_j+\frac12$ and~$h_{ij}$ does not depend on~$i$ and~$j$ in
the three uniform cases, the Hamiltonian~\eqref{Hgen} differs from~$\cH/2$ by a constant and an
irrelevant term proportional to~$\bS^2$.

In fact, $P_{ij}$ is still defined by the previous formula when the internal spin space
is~$m$-dimensional (where now~$s_i=-(m-1)/2,-(m-1)/2+1,\dots,(m-1)/2$). In this case the exchange
operators~$P_{ij}$ are related to the~$\su(m)$ generators~$t^a_k$ in the fundamental
representation (where~$k$ is the site index, $a=1,\dots,m^2-1$ and
$\tr(t^a_kt^b_k)=\frac12\de_{ab}$) by
\begin{equation}\label{Pijsum}
P_{ij}=2\sum_{a=1}^{m^2-1}t^a_it^a_j+\frac1m\equiv 2\bt_i\cdot\bt_j+\frac1m\,.
\end{equation}
With this general definition, the~Hamiltonian~\eqref{HBC} can be regarded as an~$\su(m)$ spin
model. It was shown in Ref.~\onlinecite{TS15} that in this case $\cH/2$ differs from the $\su(m)$ generalization of~\eqref{Hgen}, namely
\begin{equation}\label{HMPSm}
H_{\mathrm{MPS}}=\sum_{i\ne
  j}\bigg(\frac1{|z_i-z_j|^2}+\frac1{|z_i-z_{\bj}|^2}-\frac{h_{ij}}{4(m+1)}\bigg)\bt_i\cdot \bt_j\,,
\end{equation}
by an additive constant and a trivial term proportional to $\bT^2$, where $\bT\equiv\sum_i\bt_i$
is the total~$\su(m)$ spin operator. Moreover, as explained in Ref.~\onlinecite{TS15}, the ground
state of the Hamiltonian~$H_{\mathrm{MPS}}$ is still given by the chiral correlator~\eqref{psi},
where now
\[
A^\al(u)=\ka_\al:\e^{\iu\bmu_\al\cdot\bphi(u)/\sqrt 2}:\,,\qquad 1\le\al\le m\,,
\]
$\bphi\equiv(\phi_1,\dots,\phi_{m-1})$ is a vector of chiral bosonic fields in the~$c=m-1$ free
boson boundary~CFT. In the latter equation~$\ka_\al$ denotes a Klein factor (commuting with vertex
operators and satisfying~$\{\ka_\al,\ka_\be\}=2\de_{\al\be}$) and~$\bmu_\al\in\CC^{m-1}$ is the
weight vector of the~$\al$-th internal state in the fundamental representation of~$\su(m)$.
\section{Integrable generalization}\label{sec.IG}
We shall show in this section that the $\su(m)$~model~\eqref{HBC} can be greatly generalized
(without losing its integrability) by choosing the chain sites so that $u_1,\dots,u_N$ are the $N$
roots of the Jacobi polynomial~$P^{\be-1,\be'-1}_N$, where $\be,\be'$ are two positive parameters.
In fact, the sites in the uniform cases~\eqref{unifc} are obtained when $(\be,\be')$ respectively
take the values $(1/2,1/2)$, $(3/2,3/2)$ and $(3/2,1/2)$.

In order to understand how this integrable generalization comes about, consider to begin with the 
Hamiltonian of the open Sutherland spin \emph{dynamical} model, given by~\cite{Ya95,EFGR05}
\begin{multline}\label{HBC05}
H_{\mathrm{BC}}=-\De+a\sum_{i\ne j}\big[\sin^{-2}x_{ij}^-(a-P_{ij})+\sin^{-2}x_{ij}^+(a-\widetilde P_{ij})\big]\\
+\sum_i\big[b(b-P_i)\sin^{-2}x_i+b'(b'-P_i)\cos^{-2}x_i\big]\,,
\end{multline}
where $\De=-\sum_i\pd_{x_i}^2$, $x_{ij}^\pm=x_i\pm x_j$, $a>0$, $b=\be a$, $b'=\be' a$, $\widetilde
P_{ij}=P_iP_jP_{ij}$ and $P_i$ is the spin reversal operator, defined by
\[
P_{i}\ket{\dots,s_i,\dots}=\ket{\dots,-s_i,\dots}\,.
\]
We shall consider instead a novel variant of this model, obtained by replacing the spin reversal
operators $P_i$ by the identity, which amounts to taking a different representation of the Weyl
group of $BC_N$ type. In this way we obtain the Hamiltonian
\begin{multline}\label{HSuth}
  H=-\De+a\sum_{i\ne j}\big(\sin^{-2}x_{ij}^-+\sin^{-2}x_{ij}^+\big)(a-P_{ij})\\
  +\sum_i\big[b(b-1)\sin^{-2}x_i+b'(b'-1)\cos^{-2}x_i\big]\,,
\end{multline}
to which one can associate the auxiliary \emph{scalar} operator
\begin{multline*}
  H'=-\De+a\sum_{i\ne j}\big[\sin^{-2}x_{ij}^-(a-K_{ij})
  +\sin^{-2}x_{ij}^+(a-\tK_{ij})\big]\\
  +\sum_i\big[b\sin^{-2}x_i(b-K_i) +b'\cos^{-2}x_i(b'-K_i)\big]\,.
\end{multline*}
In the latter equation the operators~$K_{ij}$ and $K_i$ act on a scalar function
as
\begin{align*}
&K_{ij}f(\dots,x_i,\dots, x_j,\dots)=f(\dots,x_j,\dots,x_i,\dots)\,,\\
&K_if(\dots,x_i,\dots)=f(\dots,-x_i,\dots)\,,
\end{align*}
and~$\tK_{ij}=K_{ij}K_iK_j$. It was shown in Ref.~\onlinecite{EFGR05} that~$H'$ commutes with the
family of (commuting) $BC_N$-type dynamical Dunkl operators $J_k=\iu\,\pd_{x_k}+2a\ms d_k$
($k=1,\dots,N$), where
\begin{multline}\label{Dunkl}
d_k=\frac12\sum_{l(\ne k)}\Big[(1-\iu\cot
x_{kl}^-)\,K_{kl}+(1-\iu\cot x_{kl}^+)\,\tK_{kl}\Big]\\
-\sum_{l<k}K_{kl}+\frac12\big[\be(1-\iu\cot x_k)+\be'(1+\iu\tan x_k)\big]K_k\,.
\end{multline}
Equating to zero the coefficient of~$a^2$ in the commutator of $H'$ with~$J_k$ we easily
arrive at the relation
\[
[h'(\bx),d_k]=\frac\iu8\pdf{U}{x_k}\,,
\]
where~$\bx=(x_1,\dots,x_N)$ and
\begin{align*}
&U(\bx)=\sum_{i\ne j}\Big(\sin^{-2}x_{ij}^-
  +\sin^{-2}x_{ij}^+\Big)\\
  &\hphantom{U(\bx)=\sum}+\sum_i\Big(\be^2\sin^{-2}x_i+\be'^2\cos^{-2}x_i\Big)\,,\\
&h'(\bx)=\frac14\sum_{i\ne j}\Big[\sin^{-2}x_{ij}^-(1-K_{ij})
  +\sin^{-2}x_{ij}^+(1-\tK_{ij})\Big]\\
  &\hphantom{\cH(\bx)=\sum}+\frac14\sum_i\Big(\be\sin^{-2}x_i +\be'\cos^{-2}x_i\Big)(1-K_i)\,.
\end{align*}
It should be noted that the spin chain Hamiltonian~$\cH$ in Eq.~\eqref{HBC} coincides with the
operator~$-h(\bxi)$, where $h(\bx)$ is obtained from~$h'(\bx)$ by the formal replacements
$(K_{ij},K_i)\mapsto(P_{ij},1)$. Following the approach of Refs.~\onlinecite{BPS95}
and~\onlinecite{EFGR05}, the integrability condition for the Hamiltonian~$\cH$ (with chain sites
not yet determined) is the vanishing of the commutator~$[h'(\bx),d_k]$ on the chain sites~$\bxi$.
Thus a Hamiltonian of the form~\eqref{HBC} is integrable provided that its lattice sites satisfy
the system of equations
\[
\pdf{U}{x_k}(\bxi)=0\,,\qquad k=1,\dots,N\,.
\]
As is shown in
Ref.~\onlinecite{CS02}, when the parameters $\be$, $\be'$ are both positive the latter system has
essentially a unique solution determined by the conditions
\begin{equation}\label{JacobiP}
  P_N^{\be-1,\be'-1}(\cos 2\xi_j)=0\,,\qquad j=1,\dots,N\,.
\end{equation}
This establishes the integrability of the spin chain~\eqref{HBC} with sites satisfying
Eq.~\eqref{JacobiP} for arbitrary (positive) values of $\be,\be'$. In fact, it should be noted
that the parameters~$(\be',\be)$ and~$(\be,\be')$ give rise to the \emph{same}
Hamiltonian~\eqref{HBC}. Indeed, from the equality~$P^{\be,\be'}(u)=P^{\be',\be}(-u)$ it follows
that the angles~$\xi_j'$ determined by the parameters~$(\be',\be)$ are related to the~$\xi_j$'s in
Eq.~\eqref{JacobiP} by~$2\xi_j'=\pi-2\xi_{N-j+1}$, and~\eqref{HBC} is clearly invariant under the
transformation~$\xi_j\mapsto\xi_j'$. Thus we could restrict ourselves, without loss of generality,
to the case~$\be\ge\be'$. It can also be shown that the only models of the form~\eqref{HBC} with
uniformly spaced sites are precisely the three uniform cases~\eqref{unifc}.

We have remarked in Section~\ref{sec.SA} that the $\su(m)$~Hamiltonian~\eqref{HBC} essentially
reduces to~\eqref{HMPSm} for the three uniform cases~$(\be,\be')=(1/2,1/2),(3/2,3/2),(3/2,1/2)$.
It is natural to enquire whether this also holds for arbitrary positive values of~$\be$
and~$\be'$. To answer this question, we note that substituting $u_j=\cos(2\xi_j)$ ($j=1,\dots,N$)
in the system satisfied by the zeros of the Jacobi polynomial~$P_N^{\be-1,\be'-1}(u)$ listed in
Eq.~(5.2a) of Ref.~\onlinecite{ABCOP79} one obtains the equations
\begin{equation}\label{cotid}
\sum_{k(\ne j)}\big(\cot\xi_{jk}^-+\cot\xi_{jk}^+\big)=(\be'-\be)\cot
\xi_j-2\be'\cot(2\xi_j)\,,
\end{equation}
$j=1,\dots,N$\,. The relations $\iu w_{jk}=\cot\xi_{jk}^-$, $\iu w_{j\bar k}=\cot\xi_{jk}^+$ and
Eqs.~\eqref{cotid} then yield
\[
  \iu c_j=(2\be'-1)\cot(2\xi_j)+(\be-\be')\cot \xi_j\,,
\]
from which one readily obtains
\begin{equation}\label{hij}
h_{ij}\equiv (c_i-c_j)w_{ij}+(c_i+c_j)w_{i\bar\jmath}=2(\be+\be'-1)\,,
\end{equation}
in agreement with the result for the special values of $\be$ and $\be'$ mentioned above. Since the
right-hand side of~\eqref{hij} is independent of $i$ and $j$, from Eq.~\eqref{Pijsum} it again
follows that the general $\mathrm{su}(m)$ model~\eqref{HBC} with sites satisfying~\eqref{JacobiP}
with \emph{arbitrary} (positive) $\be$ and~$\be'$ is equivalent to the Hamiltonian~\eqref{HMPSm}
(i.e., $\cH=2\cH_{\mathrm{MPS}}+A\bT^2+B$ for suitable constants~$A,B$). In particular, this shows
that the ground state of the general model~\eqref{HBC} coincides with a correlator of the~$c=m-1$
free boson boundary~CFT, thus generalizing the results in Ref.~\onlinecite{TS15} for the three
uniform cases. For instance, in the~$\su(2)$ case this observation implies that the linear system
for the two-point correlation functions of the model~\eqref{Hgen} deduced in
Ref.~\onlinecite{TS15} also holds for the more general integrable model~\eqref{HBC} with sites
determined by the conditions~\eqref{JacobiP}.

We shall end this section with a brief discussion of the distribution of the sites of the general
model~\eqref{HBC} with sites~\eqref{JacobiP}. To begin with, it is shown in Ref.~\onlinecite{KA99}
that when~$N\to\infty$ the density of the zeros~$u_j$ of the Jacobi
polynomial~$P_N^{\be-1,\be'-1}$ in the interval~$[-1,1]$ approaches the continuous distribution
\[
\rho(u)=\frac1{\pi\sqrt{1-u^2}}\,.
\]
This result easily implies that the chain sites $z_j=\e^{2\iu\xi_j}$ become uniformly distributed
on the upper unit circle as~$N$ tends to infinity. In fact, according to a classical result of
Szeg\H o\cite{Sz75}, when $N\gg1$ the angles $2\xi_j$ satisfy
\[
2\xi_j=\frac{j\pi}{N}+\Or(N^{-1}),
\]
where $\Or(N^{-1})$ is uniformly bounded in $j\le N$. (Note that this property clearly holds in
the uniform cases listed in Eq.~\eqref{unifc}.) In particular, when $1/2\le\be,\be'\le3/2$ we have
the more precise estimates~\cite{Sz75}
\[
\frac{(2j-1)\pi}{2N+1}\le 2\xi_j\le\frac{2j\pi}{2N+1}\,,
\]
with equality if and only if $(\be,\be')=(1/2,3/2)$ (in the first inequality) or $(3/2,1/2)$ (in
the second one). In fact, both of these special cases give rise to the same
Hamiltonian~\eqref{HBC} ---the third uniform model in Eq.~\eqref{unifc}---, on account of the
remark following Eq.~\eqref{JacobiP}.

\section{Twisted Yangian symmetry}\label{sec.tYs}
It was shown in Ref.~\onlinecite{BPS95} that the~spin $1/2$ model~\eqref{HBC} with $(\be,\be')$ in
the three uniform cases mentioned above possesses a monodromy matrix $T(u)$ which satisfies the
reflection equation~\cite{Sk88}. More precisely, the monodromy matrix is given by~\cite{BPS95}
\[
T(u)=\bigg(1+\frac{\be+\be'}{2u}\bigg)\widetilde T(u),
\]
where
\[
\widetilde T(u)=\pi\left[\prod_{i=1}^N\bigg(1+\frac{P_{0i}}{u-d_i}\bigg)
  \prod_{i=N}^1\bigg(1+\frac{P_{0i}}{u+d_i}\bigg)\right]
\]
and~$d_i$ is defined in Eq.~\eqref{Dunkl}. In the latter formula the index~$0$ labels an auxiliary
$m$-dimensional internal space, and the projection operator~$\pi$ is defined by~$\pi(x_j)=\xi_j$
and
\[
\pi(K_{i_1j_1}\cdots K_{i_rj_r}K_{l_1}\cdots K_{l_{s}})=P_{i_rj _r}\cdots P_{i_1j_1}\,.
\]
This monodromy matrix is actually valid for~$\su(m)$ spin~\cite{TS15} and arbitrary (positive)
values of the parameters~$(\be,\be')$, since it only depends on the expression of the Hamiltonian
in terms of permutation operators~$P_{ij}$ and the integrability conditions~\eqref{JacobiP}. Thus
the general $\su(m)$ model~\eqref{HBC}, with sites satisfying Eq.~\eqref{JacobiP}, possesses
twisted Yangian symmetry. Since, by construction, $[\cH,T(u)]=0$ for all $u$, the coefficients
of~$1/u$ in the Laurent expansion of~$T(u)$ form a family of conserved charges for the
Hamiltonian~$\cH$ of the open Haldane--Shastry chain. In fact, since the term $1+(\be+\be')/2u$ is
a scalar, the conserved charges can be more directly obtained by expanding the remaining term
in~$T(u)$. In other words, if
\[
\widetilde T(u)=1+\sum_{n=1}^\infty\frac{J_n}{u^n}
\]
then~$[\cH,J_n]=0$ for all $n\ge1$. To begin with, from the expansions
\[
\frac1{u\mp d_i}=\frac1u\pm\frac{d_i}{u^2}+\frac{d_i^2}{u^3}+\Or(u^{-4})
\]
and the identity
\begin{equation}
  P_{0i}=\frac1m+2\sum_at_0^at_i^a
  \label{P0it}
\end{equation}
we easily obtain~\footnote{Unless otherwise specified, here and in what follows sums over the
  $\su(m)$ indices~$a,b,c$ range from $1$ to $m^2-1$.}
\[
J_1=2\sum_iP_{0i}=\frac{2N}m+4\sum_{a}t_0^a\sum_it_i^a\,.
\]
Thus the first-order conserved charges are the total~$\su(m)$ generators~$T^a$, $1\le a\le m^2-1$.
Similarly, expanding~$\widetilde T(u)$ to second order we obtain
\begin{align*}
J_2&=\left(\sum_{i<j}+\sum_{i>j}+\sum_{i,j}\right)P_{0i}P_{0j}=N+2\sum_{i\ne j}P_{0i}P_{0j}\\
&=N+2\sum_{i\ne j}P_{ij}P_{0i}\,,
\end{align*}
which yields the second-order conserved charges
\[
J_2^0=\sum_{i\ne j}P_{ij}\,,\qquad
J_2^a=\sum_{i\ne j}P_{ij}t^a_i\,.
\]
Both of these charges, however, are trivial, in the sense
that they are polynomial functions of the global $\su(m)$ generators~$T^a$. Indeed,
\begin{align*}
  J_2^0&=\frac Nm\,(N-1)+2\sum_{i\ne j}\sum_at^a_it^a_j\\
  &=\frac Nm\,(N-1)+2\sum_a(T^a)^2
  -2\sum_i\sum_a(t_i^a)^2\\
       &=\frac Nm\,(N-1)+2\sum_a(T^a)^2-\frac Nm\,(m^2-1)\\
       &=2\sum_a(T^a)^2+\frac Nm\,(N-m^2)\,.
\end{align*}
The conserved charges~$J^a_2$ can be simplified using the identity~\eqref{Pijsum} and the
relations
\begin{equation}\label{tatb}
t^at^b=\frac1{2m}\de_{ab}+\frac12\sum_{c}(d_{abc}+\iu f_{abc})t^c
\end{equation}
satisfied by the $\su(m)$ generators in the fundamental representation. Here $d_{abc}$ and~$f_{abc}$ are respectively totally symmetric and totally
antisymmetric in $a,b,c$, with
\begin{equation}\label{dabc}
\sum_{b}d_{abb}=0\,,\qquad \sum_{b,c}d_{abc}d_{bcr}=\frac{m^2-4}m\,\de_{ar}
\end{equation}
(see, e.g., Ref.~\onlinecite{MSW68}). After a straightforward calculation one obtains
\[
J_2^a=\bigg(\frac{2N}m-\frac m2\bigg)T^a+\sum_{b,c}d_{abc}T^bT^c\,.
\]

Similarly, expanding~$\widetilde T(u)$ to third order in $1/u$ and simplifying slightly the result
we arrive at the expression
\begin{align*}
  \frac{J_3}2&=\sum_iP_{0i}\pi(d_i^2)+\sum_{i<j}(P_{0i}P_{0j}-P_{0j}P_{0i})\pi(d_i)\\
  &\hphantom{{}={}}+\frac12\bigg(\sum_{i<j<k}+\sum_{i<j,k}\bigg)(P_{0i}P_{0j}P_{0k}+P_{0k}P_{0j}P_{0i})\\
             &=\sum_iP_{0i}\pi(d_i^2)+\sum_{i<j}(P_{0j}-P_{0i})P_{ij}\pi(d_i)\\
             &\hphantom{{}={}}+\bigg(\sum_{i<j<k}+\sum_{k<i<j}\bigg)(P_{0k}P_{ik}P_{jk}+P_{0i}P_{ik}P_{ij})\\
  &\hphantom{{}={}}+\sum_{i<j}(P_{0i}+P_{ij})\,.
\end{align*}
Using again Eq.~\eqref{P0it} to identify the coefficients of~$t^0_0\equiv\id$ and~$t^a_0$ in the
previous formula, and dropping the trivially conserved term~$\sum_{i<j}P_{ij}$, we obtain the
third-order conserved charges
\begin{align*}
  J_3^0&=\sum_i\pi(d_i^2)+\bigg(\sum_{i<j<k}+\sum_{k<i<j}\bigg)P_{ik}(P_{jk}+P_{ij})\,,\\
  J_3^a&=\sum_it^a_i\pi(d_i^2)+\sum_{i<j}(t^a_{j}-t^a_{i})P_{ij}\pi(d_i)+\sum_{i}(N-i)t^a_{i}\\
  &\hphantom{{}={}}+\bigg(\sum_{i<j<k}+\sum_{k<i<j}\bigg)(t^a_{k}P_{ik}P_{jk}+t^a_{i}P_{ik}P_{ij})\,.
\end{align*}
The first of these charges is trivial. Indeed, note first of all that from the coefficient of
$a^2$ in the identity $H'=\sum_iJ_i^2$, where~$J_k=\iu\pd_{x_k}+2a d_k$
(cf.~Ref.~\onlinecite{EFGR05}), it follows that $\sum_i\pi(d_i^2)=U(\bxi)/4$ is a constant.
Secondly, from the permutation group identities $P_{ik}P_{jk}=P_{jk}P_{ij}=P_{ij}P_{ik}$ (valid
when $i,j,k$ are all distinct) we easily obtain
\begin{multline*}
  \bigg(\sum_{i<j<k}+\sum_{k<i<j}\bigg)P_{ik}(P_{jk}+P_{ij})\\
 =
    \frac13\sideset{}{'}\sum_{i,j,k}P_{ik}(P_{jk}+P_{ij})
  =\frac23\sideset{}{'}\sum_{i,j,k}P_{ik}P_{jk}\,,
\end{multline*}
where the last sum can be shown to be trivially conserved by repeated application
of~Eqs.~\eqref{tatb}-\eqref{dabc}. As to the remaining third-order charges~$J_3^a$, a long but
straightforward calculation yields
\begin{multline*}
  -4J_3^a=\sum_{i\ne
  j}(w_{ij}-w_{i\bar\jmath})^2t^a_i+\sum_i\big[(\be-\be')w_{i0}+2\be'w_{i\bar\imath}\big]^2t^a_i\\
-\sideset{}{'}\sum_{i,j,k}(w_{ij}-w_{i\bar\jmath})(w_{jk}+w_{j\bar k})t^a_iP_{ik}P_{ij}\,.
\end{multline*}
The first two terms in the latter expression can be simplified with the help of Eq.~\eqref{sum}
and the identity $2w_{i0}w_{i\bar\imath}=1+w_{i0}^2$. Dividing the resulting expression
for~$-4J_3^a$ by the the nonzero coefficient~$8\be'(\be'+1)/3$ and dropping a trivially conserved
term proportional to $T^a$ we finally obtain the equivalent non-trivial conserved charges
\begin{multline*}
Q^a=\sum_i\big(w_{i\bar\imath}^2+\ga_1w_{i0}^2\big)t^a_i\\
-\ga_2\sideset{}{'}\sum_{i,j,k}(w_{ij}-w_{i\bar\jmath})(w_{jk}+w_{j\bar k})t^a_iP_{ik}P_{ij}\,,
\end{multline*}
where the coefficients~$\ga_{1,2}$ are
given by
\[
\ga_1=\frac{(\be-\be')(1+\be+\be')}{4\be'(\be'+1)}\,,\qquad
\ga_2=\frac{3}{8\be'(\be'+1)}\,.
\]
In particular, for the values of~$(\be,\be')$ corresponding to the three uniform cases the
previous expression is in agreement~\footnote{Actually, Eq.~(8) of Ref.~\onlinecite{TS15} contains
  a minor typo, i.e., the term~$w_{ik}+w_{i\bar k}$ in the last sum should
  be~$w_{jk}+w_{j\bar k}$} with Eq.~(8) in Ref.~\onlinecite{TS15}.

\section{Partition function}\label{sec.PF}
We shall next evaluate in closed form the partition function of the $\su(m)$~spin
chain~\eqref{HBC}. The key idea in this respect is to exploit the connection between the latter
model and the $\su(m)$ spin Sutherland model~\eqref{HSuth} by means of the so-called freezing
trick~\cite{Po93,SS93,Po94}. More precisely, when~$a\to\infty$ the Hamiltonians $\cH$ and $H$ are
related by $H\simeq H_{\mathrm{sc}}-4a\,\cH$, where $H_{\mathrm sc}$ is the Hamiltonian of the
scalar Sutherland model obtained from~$H$ by replacing $P_{ij}$ by $1$. From the latter relation
it follows that the partition functions~$Z$, $Z_{\mathrm{sc}}$ and $\cZ$ of the Hamiltonians~$H$,
$H_{\mathrm{sc}}$, $\cH$, respectively, are related by~\cite{Po94}
\begin{equation}\label{FTr}
\cZ(T)=\lim_{a\to\infty}\frac{Z(-4aT)}{Z_{\mathrm sc}(-4aT)}\,.
\end{equation}
Thus the partition function~$\cZ$ can
be evaluated from the spectra of the Hamiltonians~$H$ and $H_{\mathrm sc}$, which in turn can be
derived from that of the auxiliary operator~$H'$ following the approach
of~Ref.~\onlinecite{EFGR05}.

The spectrum of~$H'$ can be computed by noting that it acts triangularly on the (non-orthonormal)
basis
\begin{equation}\label{S.phin}
\phi_{\bn}(\bx)=\phi(\bx)\,\e^{2\iu\,\bn\cdot\bx}\,\quad \bn=(n_1,\dots,n_N)\in\ZZ^N\,,
\end{equation}
where
\mbox{$\phi(\bx)=\prod\limits_{i<j}|\sin x_{ij}^-\, \sin x_{ij}^+|^a\cdot \prod\limits_i|\sin x_i|^{b}
  |\cos x_i|^{b'}$}.
More precisely, we introduce a partial ordering $\prec$ in the basis~\eqref{S.phin} as follows.
Given a multiindex $\bn\in\ZZ^N$, we define the nonnegative and nonincreasing multiindex $[\bn]$
by $ [\bn]=(|n_{i_1}|,\dots,|n_{i_N}|)$, where $|n_{i_1}|\ge\dots\ge|n_{i_N}|$\,. If
$\bn,\bn'\in[\ZZ^N]$ are two such multiindices, we shall say that $\bn\prec \bn'$ if
$n_1-n_1'=\dots=n_{i-1}-n_{i-1}'=0$ and $n_i <n_i'$. For arbitrary $\bn,\bn'\in\ZZ^N$, we shall
say that $\bn\prec \bn'$ or $\phi_{\bn}\prec\phi_{\bn'}$ provided that $[\bn]\prec[\bn']$. With
the help of this partial ordering, it can be shown that
\begin{equation}\label{S.Hpphin}
H'\phi_{\bn}=E_{\bn}\phi_\bn + \sum_{\bn'\prec\bn}c_{\bn',\bn}\phi_{\bn'}\,,
\end{equation}
where the eigenvalue $E_{\bn}$ is given by~\cite{EFGR05}
\begin{equation}\label{Ebn}
E_{\bn}=\sum_i\big(2[\bn]_i+b+b'+2a(N-i)\big)^2\,.
\end{equation}
From the basis~\eqref{S.phin} one can construct a set of spin wavefunctions spanning the Hilbert
space of the Hamiltonian $H$ by applying the operator $\La$ which projects onto states symmetric
under particle permutations and reflections of the spatial coordinates, determined by the
relations $K_{ij}\La=P_{ij}\La$, $K_i\La=\La$. In this way we obtain the set of spin wavefunctions
\begin{equation}\label{S.psis}
  \psi_{\bn,\bs}(\bx)=\phi(\bx)\La\big(\e^{2\iu\,\bn\cdot\bx}\ket\bs\big),\quad \ket\bs\equiv \ket{s_1,\dots,s_N}\,.
\end{equation}
It is clear that these wavefunctions are not linearly independent. However, using the properties
of the projector~$\La$ one can easily extract from the set of wavefunctions~\eqref{S.psis} a
(non-orthonormal) basis~$\cB$ by suitably restricting the quantum numbers $\bn$ and~$\bs$. A
convenient way of achieving this end is by imposing the following conditions:
\begin{enumerate}[i)]
\item $n_1\ge n_2\ge\cdots\ge n_N\ge0$, i.e, $\bn\in[\ZZ^N]$\,.
\item If $n_i=n_j$, then  $s_i\ge s_j$\,.
\end{enumerate}
From Eq.~\eqref{S.Hpphin} and the relation $H'\La=H\La$ one can easily check that the action of
$H$ on the basis~$\cB$ is given by $H\psi_{\bn,\bs}=E_{\bn}\psi_{\bn,\bs} + \text{l.o.t}$,
where~l.o.t.\ denotes a linear combination of basis functions with quantum numbers~$(\bn',\bs')$
satisfying~$\bn'\prec\bn$. Thus the Hamiltonian~$H$ is again upper triangular in the basis~$\cB$,
partially ordered according to the prescription~$\psi_{\bn,\bs}\prec\psi_{\bn',\bs'}$ if
$\bn\prec\bn'$. Its eigenvalues~$E_{\bn,\bs}=E_{\bn}$ are given by Eq.~\eqref{Ebn} or, taking into
account that~$[\bn]=\bn$ by condition~i),
\[
E_{\bn,\bs}=4\sum_i\big(n_i+a(\bar\be+N-i)\big)^2\,,\qquad\bar\be\equiv\frac{\be+\be'}2\,.
\]
Writing $ \bn=(\underbrace{\nu_1,\dots,\nu_1}_{k_1},\dots,\underbrace{\nu_r,\dots,\nu_r}_{k_r}) $,
it is clear that the energies $E_{\bn,\bs}$ have an intrinsic degeneracy
$d(\bn)=\prod_{i=1}^r\binom{m+k_i-1}{k_i}$ equal to the number of basis states~$\ket\bs$
compatible with condition~ii) above. From the expansion
\[
\frac{E_{\bn,\bs}}{4a}=\frac{a E_0}4+2\sum_i n_i(\bar\be+N-i)+\Or(1/a),
\]
where $a^2E_0$ is the ground state energy of $H$, we obtain
\[
\lim_{a\to\infty}\big[q^{\frac{aE_0}{4}}Z(-4aT)\big]=\sum_{n_1\ge \cdots\ge
  n_N\ge0}d(\bn)\,q^{2\sum_in_i(i-N-\bar\be)}\,.
\]
Here
\[
q\equiv\e^{-1/(k_{\mathrm B}T)}\,,
\]
where $T$ is the temperature and~$k_{\mathrm B}$ is Boltzmann's constant. The sum in the exponent
of the RHS can be expressed in terms of~$\nu_i$ and $k_i$ as
\begin{equation}\label{S.sumi}
\sum_{i=1}^r\nu_ik_i(2N_i+1-2\bar\be-2N-k_i)\,,
\end{equation}
where~$N_i\equiv\sum_{j=1}^ik_i$. Proceeding as in Ref.~\onlinecite{EFGR05}, we introduce the new
variables $l_i=\nu_i-\nu_{i+1}>0$ ($i=1,\dots,r-1$) and $l_r=\nu_r\ge0$. After a straightforward
calculation one can then rewrite the sum~\eqref{S.sumi} as~$\sum_{j=1}^rl_j\cE(N_j)$, where
\begin{equation}\label{disprel}
  \cE(j)=j(j+1-2\bar\be-2N)\,.
\end{equation}
Denoting by~$\cP_N$ the set of all partitions of the integer $N$ (with order taken into account)
we get the compact formula
\begin{align*}
  \lim_{a\to\infty}\big[&q^{\frac{aE_0}{4}}Z(-4aT)\big]\\
  &=\sum_{\bk\in\cP_N}\prod_{i=1}^r{\textstyle\binom{m+k_i-1}{k_i}}
                          \sum_{\substack{l_1,\dots,l_{r-1}>0\\l_r\ge0}}
  \prod_{j=1}^rq^{l_j\cE(N_j)}\\
                        &=\frac{1}{1-q^{N}}\sum_{\bk\in\cP_N}\prod_{i=1}^r
                          {\textstyle\binom{m+k_i-1}{k_i}}\cdot
                          \prod_{j=1}^{r-1}\frac{q^{\cE(N_j)}}{1-q^{\cE(N_j)}}\,,
\end{align*}
where $\bk=(k_1,\dots,k_r)$ and we have taken into account that~$N_r=N$. On the other hand, the
partition function of the scalar Sutherland model~$H_{\mathrm{sc}}$ was computed in
Ref.~\onlinecite{EFGR05}, with the result
\[
\lim_{a\to\infty}\big[q^{\frac{aE_0}{4}}Z_{\mathrm{sc}}(-4aT)\big]=\prod_{i=1}^N\big(1-q^{\cE(i)}\big)^{-1}\,.
\]
From the last two equations and the freezing trick relation~\eqref{FTr} one finally arrives at the
following explicit formula for the partition function of the chain~\eqref{HBC}:
\begin{equation}
  \label{cZfinal}
  \cZ(T) = \sum_{\bk\in\cP_N}\prod_{i=1}^r{\textstyle\binom{m+k_i-1}{k_i}}\cdot q^{\sum_{i=1}^{r-1}\cE(N_i)}
  \prod_{j=1}^{N-r}\big(1-q^{\cE(N_j')}\big)\,,
\end{equation}
where $N_i=\sum_{j=1}^ik_i$\,,
\[
\{N_1',\dots,N_{N-r}'\}=\{1,\dots,N-1\}\setminus\{N_1,\dots,N_{r-1}\}\,,
\]
and the \emph{dispersion relation}~$\cE$ is given by~Eq.~\eqref{disprel}.

Interestingly, the structure of the partition function~\eqref{cZfinal} is the same as that of the
original \emph{closed} Haldane--Shastry chain~\cite{FG05}, albeit with a different dispersion
relation depending on the single free (positive) parameter~$\bar\be\equiv(\be+\be')/2$. It was
shown in Ref.~\onlinecite{BBH10} that a partition function of the form~\eqref{cZfinal} coincides
with that of a related vertex model regardless of the functional form of the dispersion relation.
More precisely, consider a one-dimensional classical vertex model consisting of $N+1$ vertices
connected by $N$ intermediate bonds. Any possible state for this vertex model can be represented
by a path configuration given by a vector $\bs=(s_1,\dots,s_N)$, where $s_i\in\{1,2,\dots,m\}$
denotes the spin state of the $i$-th bond. The energy function associated with this spin path
configuration $\bs$ is defined as
\begin{equation}
  E(\bs) = \sum_{j=1}^{N-1}\cE(j)\,\theta(s_j-s_{j+1})\,, 
\label{vE}
\end{equation}
where $\theta$ is Heaviside's step function given by~$\th(x)=0$ for $x\le0$ and $\th(x)=1$
otherwise. As shown in Ref.~\onlinecite{BBH10}, the partition function of this vertex model is
given by Eq.~\eqref{cZfinal}. An important consequence of this fact is that the spectrum of the
$\su(m)$ model~\eqref{HBC} with sites satisfying~\eqref{JacobiP}, including the degeneracy of each
level, is given by Eq.~\eqref{vE}, where~$\bs$ runs over all possible $m^N$ spin configurations.
In fact, it is well-known that the spectrum of many Yangian-invariant spin models, including the
original Haldane--Shastry chain, is given by a formula of the type~\eqref{vE} with a suitable
model-dependent dispersion relation $\cE(j)$. This suggests that the chain~\eqref{HBC} may also be
invariant under the (untwisted) Yangian $Y(\mathrm{gl}(m))$ for arbitrary values of $\be$ and
$\be'$.

\section{Ground state and spectrum}\label{sec.GSS}
From Eq.~\eqref{vE} it follows that the energy levels of the chain~\eqref{HBC} can be computed
from the formula
\begin{equation}\label{motifs}
E_{\bde}=\sum_{j=1}^{N-1}\cE(j)\de_j\,,
\end{equation}
where~each $\de_j$ is either zero or one, and the vector~$\bde=(\de_1,\dots,\de_{N-1})$, which is
called a \emph{motif}, cannot contain a sequence of~$m$ or more consecutive $1$'s. In fact,
Eq.~\eqref{motifs} is the counterpart of Haldane's formula for the energies of the closed
(antiferromagnetic) $\su(m)$ Haldane--Shastry chain in terms of motifs~\cite{HHTBP92,Ha93}, for
which the dispersion relation is given by $\cE(j)=j(j-N)$. Note, however, that Eq.~\eqref{motifs},
unlike~\eqref{vE}, does not convey complete information on the degeneracy of each level. In this
section we shall use Eq.~\eqref{motifs} for the spectrum in terms of Haldane's motifs to compute
the ground state energy and deduce an alternative expression for the distinct energy levels in
terms of rapidities. This expression generalizes to arbitrary positive values of~$\be$ and~$\be'$
the formula conjectured in Ref.~\onlinecite{TS15} for the three uniform cases.

We shall begin by computing the ground state energy. In the first place, since $\cE(j)<0$ for
$j=1,\dots,N-1$ and the only restriction on the motif~$\bde\equiv(\de_1,\dots,\de_{N-1})$ is that
it cannot contain more than $m-1$ consecutive~$1$'s, it is clear that when $m\ge N$ the ground
state energy~$E_0$ is obtained from the motif~$\bde=(1^{N-1})$. Hence in this case we simply have
\begin{equation}\label{E0mgeN}
E_0=\sum_{j=1}^{N-1}\cE(j)=-\frac13\,N(N-1)(3\bar\be+2N-1)\,.
\end{equation}
On the other hand, when~$m<N$ the motif that yields the minimum energy
is~$(1^{N_0-1},0,1^{m-1},\dots,0,1^{m-1})$, where $N_0=N\bmod m$ and the~$0$'s are in the
positions $N-jm$ with~$j=1,\dots,\lfloor (N-1)/m\rfloor$. Here $\lfloor x\rfloor$ denotes the
integer part of the real number~$x$, and it is understood that if~$N_0=0$ the first
sequence~$(1^{N_0-1},0)$ is missing. Inserting the latter motif~into Eq.~\eqref{motifs} we obtain
\begin{align*}
E_0&=\sum_{j=1}^{N-1}\cE(j)-\sum_{j=1}^{\lfloor (N-1)/m\rfloor}\cE(N-jm)\\
&=\sum_{j=1}^{N-1}\cE(j)-\sum_{j=1}^{(N-N_0)/m}\cE(N-jm)\,.
\end{align*}
Indeed, $\lfloor N/m\rfloor=(N-N_0)/m=\lfloor(N-1)/m\rfloor+\de_{0,N_0}$, and the spurious term
with $j=N/m$ in the last sum when~$N_0=0$ (i.e., when $N$ is divisible by~$m$) is of no
consequence since~$\cE(0)=0$. Evaluating the last sum in the previous equation we easily find the
following explicit formula for the ground state energy when~$m<N$:
\begin{multline*}
E_0=-\frac{m-1}{6m}\,N\big[4N^2+3(2\bar\be-1)N+m\big]\\
+\frac{N_0(m-N_0)}{6m}\,\big[3(2\bar\be+2N-1)+m-2N_0\big]\,.
\end{multline*}

We shall next express Eq.~\eqref{motifs} in terms of rapidities and compare the resulting formula
with that conjectured in Ref.~\onlinecite{TS15} for the three uniform cases. More precisely, given
a motif~$\bde=(\de_1,\dots,\de_{N-1})$ its corresponding \emph{rapidities} are the positions $r_i$
($i=1,\dots,n$) of its nonzero (i.e., $1$) components, in terms of which the motif's energy is
given by
\begin{equation}\label{Eraps}
E_{\bde}=\sum_{i=1}^n\cE(r_i)\,.
\end{equation}
By Haldane's restriction on the~$\su(m)$ motifs, there cannot be more than $m-1$ consecutive
rapidities. Moreover, the maximum number of rapidities in a motif, $n_{\mathrm{max}}$, is equal to
the number of~$1$'s in the ground state motif, namely $\lfloor (m-1)N/m\rfloor$. Since
\[
\cE(j)=\bigg(\frac\nu2-j\bigg)^2-\frac{\nu^2}4\,,\qquad \nu\equiv2(\bar\be +N)-1\,,
\]
Eq.~\eqref{Eraps} can be rewritten as
\begin{equation}
  \label{Eraps2}
  E_{\bde}=\sum_{i=1}^n\bigg(\rho_i^2-\frac{\nu^2}4\bigg)\,,
\end{equation}
with
\begin{equation}
  \label{rhoi}
  \rho_i=\frac\nu2-r_i=\bar\be +N-\frac12-r_i>0\,.
\end{equation}
Note that, unlike the rapidities~$r_i$, the numbers~$\rho_i$ are generally not integers. From the
latter equation it follows that the set of distinct energy levels of the $\su(m)$ open
Haldane--Shastry chain can be generated from Eq.~\eqref{Eraps2} with the following two rules:
\begin{enumerate}[i)]
\item The $n$ numbers~$\rho_i$ belong to the
  set~$\{\bar\be+1/2,\bar\be+3/2,\dots,\bar\be+N-3/2\}$, with
  $n=0,1,\dots,\lfloor(m-1)N/m\rfloor$.
\item There can be no more than $m-1$ consecutive~$\rho_i$'s.
\end{enumerate}
Note that the second rule is a direct consequence of Haldane's restriction on the $\su(m)$ motifs,
and that by definition~$\rho_i$ and~$\rho_j$ are consecutive if their difference is equal to~$1$.
The latter description of the set of distinct energies can be easily adapted to the alternative
Hamiltonian
\begin{equation}
  \label{cHt}
  \widetilde\cH=\sum_{i\ne j}\bigg(\frac1{|z_i-z_j|^2}+\frac1{|z_i-z_{\bj}|^2}\bigg)\,\bt_i\cdot \bt_j
\end{equation}
considered in Ref.~\onlinecite{TS15} in the three uniform cases~\eqref{unifc}. To begin with, note
that from the identity~\eqref{Pijsum} it follows that
\[
\widetilde\cH=\frac12\,\cH+\frac12\bigg(1-\frac1m\bigg)\sum_{i\ne
  j}\bigg(\frac1{|z_i-z_j|^2}+\frac1{|z_i-z_{\bj}|^2}\bigg)\,.
\]
The last term coincides with the constant energy
\[
\widetilde E_0\equiv\frac{m-1}{8m}\,\Big[-\sum_{i\ne j}(w_{ij}^2+w_{i\bar\jmath}^2)+2N(N-1)\Big]
\]
in Ref.~\onlinecite{TS15}, on account of the identities
\begin{align*}
-w_{ij}^2&=\cot^2\xi_{ij}^-=\sin^{-2}\xi_{ij}^--1=\frac{4}{|z_i-z_j|^2}-1\,,\\
-w_{i\bar\jmath}^2&=\cot^2\xi_{ij}^+=\sin^{-2}\xi_{ij}^+-1=\frac{4}{|z_i+z_j|^2}-1\,.
\end{align*}
Thus the distinct energies of the Hamiltonian~$\widetilde\cH$ can be generated from the formula
\begin{equation}\label{tildeE}
\widetilde E=\widetilde E_0+\frac12\sum_{i=1}^n\bigg(\rho_i^2-\frac{\nu^2}4\bigg)
\end{equation}
by the two rules above for the numbers~$\rho_i$. In particular, in the three uniform
cases~$\bar\be=1/2,3/2,1$ the $\rho_i$'s belong to the sets $\{1,\dots,N-1\}$, $\{2,\dots,N\}$,
$\{3/2,5/2,\dots,N-1/2\}$ and~$\nu=2N,2(N+1),2N+1$, respectively, so that Eq.~\eqref{tildeE}
reduces to the formula conjectured in Ref.~\onlinecite{TS15}. Note, finally, that the constant
energy~$\widetilde E_0$ in the previous equation can be easily computed by noting that
\[
\widetilde E_0=\frac{m-1}{2m}\,\sum_{i\ne
  j}\bigg(\frac1{|z_i-z_j|^2}+\frac1{|z_i-z_{\bj}|^2}\bigg)=\frac{1-m}{4m}\,E_0\,,
\]
where~$E_0$ is the ground state energy of the Hamiltonian~$\cH$ when $m\ge N$, given by
Eq.~\eqref{E0mgeN}. We thus have
\[
\widetilde E_0=\frac1{12}\,\bigg(1-\frac1m\bigg)N(N-1)(3\bar\be+2N-1)\,.
\]

The fact that the spectrum of the spin chain~\eqref{HBC} is fully described by
Eqs.~\eqref{disprel} and~\eqref{vE} has several important consequences that we shall now discuss.
Indeed, it was shown in Refs.~\onlinecite{EFG10,BB12} that the level density of any quantum system
whose spectrum is of the form~\eqref{vE} with a dispersion relation~$\cE(j)$ polynomial in $j$ and
$N$ is normally distributed in the limit $N\to\infty$. Secondly, since an equation of the
form~\eqref{vE} also describes the spectrum of Yangian-invariant $\su(m)$ spin models, the
spectrum of the chain~\eqref{HBC} must be highly degenerate for \emph{all} values of~$\be$ and
$\be'$. In fact, from the polynomial character of this chain's dispersion relation it follows that
its average degeneracy should be much higher than that of a generic Yangian-invariant model (with
a non-polynomial dispersion relation). More precisely, it was shown in Ref.~\onlinecite{FG15} that
when $\cE(j)$ is a polynomial in $j$ and $N$ the number of distinct levels is at
most~$\Or\big(N^{\sum_s(s+1)r(s)}\big)$, where, for a given~$s$, $r(s)$ is the number of monomials
of the form~$N^p j^s$ in $\cE$. Moreover, when $\cE(j)$ is a polynomial with \emph{rational}
coefficients the number of distinct levels is actually $\Or(N^{k+1})$, where~$k$ is the total
degree of~$\cE(j)$ in $j$ and $N$. For the dispersion relation~\eqref{disprel} we have $r(1)=2$,
$r(2)=1$ and $k=2$, so that the number of distinct levels of the chain~\eqref{HBC} is (at most)
$\Or(N^7)$ for arbitrary $\be+\be'$ and $\Or(N^3)$ for rational $\be+\be'$. This is indeed much
lower than for a generic Yangian-invariant spin model, for which the latter number grows
exponentially~\cite{FG15} with~$N$.

\section{Conclusions and outlook}\label{sec.CO}
In this paper we have introduced an integrable generalization of the $\su(m)$ Simons--Altshuler
open chain~\cite{SA94,BPS95,TS15} depending on two arbitrary positive parameters~$\be$ and~$\be'$,
whose sites are determined by the zeros of a suitable Jacobi polynomial. Using the results in
Ref.~\onlinecite{TS15}, we have shown that this model's ground state can be obtained from the
chiral correlator of the $c=m-1$ free boson boundary CFT. We have computed the first few
nontrivial conserved charges stemming from the model's twisted Yangian symmetry, and evaluated the
chains' partition function in closed form for arbitrary values of its parameters. From the
partition function we have been able to deduce a formula for the energy spectrum in terms of
Haldane's motifs, with a dispersion relation similar to that of the original (closed)
Haldane--Shastry chain. As shown in Ref.~\onlinecite{EFG12}, this formula can be applied to derive
the chain's thermodynamical properties, which could be relevant in the context of the
single-impurity Kondo problem. Finally, it should be noted that the chain's connection to a
conformal field theory could be exploited in several different ways. For instance, in the~$\su(2)$
case it is well known that the spin correlation functions of this type of models satisfy a system
of linear equations whose coefficients depend on the chain sites in a simple way~\cite{TS15}. This
fact, which was used in the latter reference to compute the correlators in the first uniform case,
provides a promising way for evaluating the correlators in the $\su(2)$ case for arbitrary values
of the parameters~$\be$ and~$\be'$.

\section*{Acknowledgments}
The authors would like to thank H.-H. Tu for his helpful comments on a previous version of this
manuscript. This work was partially supported by Spain's MINECO under grant no.~FIS2015-63966, and
by the Universidad Complutense de Madrid and Banco Santander under grant no.~GR3/14-910556.

\appendix*
\section{Summation formula for the zeros of Jacobi polynomials}
In this appendix we shall prove the identity
\begin{align}\label{sum}
  \sum_{j(\ne i)}&(w_{ij}-w_{i\bar\jmath})^2
  =\frac13\,(\be-\be')(2-\be-\be')w_{i0}^2
  \notag\\&+\frac43\,\be'(2-\be')w_{i\bar\imath}^2
  -\frac{4N^2}3+\frac{4N}3\,(4-\be-\be')\notag\\
  &-\frac23\,(1+\be')(\be-\be')+\frac{8\be}3-4\,,
\end{align}
which is used in Section~\ref{sec.tYs} to simplify the conserved quantities of the open
Haldane--Shastry chain. To begin with, note that
\[
u_j=\Re z_j=\frac12(z_j+z_j^{-1})=\frac{z_j^2+1}{2z_j}\,,
\]
and hence
\begin{align*}
u_i-u_j&=\frac{(z_j-z_i)(1-z_iz_j)}{2z_iz_j}\,,\\
w_{ij}-w_{i\bar\jmath}&=\frac{z_i+z_j}{z_i-z_j}-\frac{z_i+z_j^{-1}}{z_i-z_j^{-1}}
=\frac{z_j^2-1}{z_j}\,(u_i-u_j)^{-1}\,.
\end{align*}
Taking into account that $z_j=\e^{2\iu\xi_j}$, so that
\[
\left(\frac{1-z_j^2}{z_j}\right)^2=\left(\frac1{z_j}-z_j\right)^2=-4\sin^2(2\xi_j)=-4(1-u_j^2)\,,
\]
we have
\begin{equation}\label{sumorig}
\sum_{j(\ne i)}(w_{ij}-w_{i\bar\jmath})^2=-4\sum_{j(\ne i)}\,\frac{1-u_j^2}{(u_i-u_j)^2}
\,.
\end{equation}
Since
\[
\frac{1-u_j^2}{(u_i-u_j)^2}=\frac{1-u_i^2}{(u_i-u_j)^2}+\frac{2u_i}{u_i-u_j}-1\,,
\]
the sum in the RHS of Eq.~\eqref{sumorig} can be evaluated using Eqs.~(5.2a)-(5.2b) in
Ref.~\onlinecite{ABCOP79}, which in our notation read
\begin{align*}
  &2(1-u_i^2)\sum_{j(\ne i)}(u_i-u_j)^{-1}
  =\be-\be'+(\be+\be')u_i\,,\\
  &12(1-u_i^2)^2\sum_{j(\ne i)}(u_i-u_j)^{-2}
    =4(N-1)(N+\be+\be')\\
  &\qquad-(\be-\be')^2
    -2(\be-\be')(4+\be+\be')u_i\\
  &\qquad-\big[4N(N+\be+\be'-1)+(\be+\be')(4+\be+\be')\big]u_i^2\,.
\end{align*}
Using the latter formulas and the identities
\begin{align*}
w_{i\bar\imath}^2&=\left(\frac{z_i^2+1}{z_i^2-1}\right)^2=-\cot^2(2\xi_i)=-\frac{u_i^2}{1-u_i^2}\,,
\\
w_{i0}^2&=\left(\frac{z_i+1}{z_i-1}\right)^2=-\cot^2\xi_i=\frac{u_i+1}{u_i-1}\,,
\end{align*}
from which it follows that
\[
\frac{u_i}{1-u_i^2}=\frac1{1-u_i}-\frac1{1-u_i^2}=\frac12\,(2w_{i\bar\imath}^2-w_{i0}^2-1)\,,
\]
we obtain Eq.~\eqref{sum} after a straightforward calculation.

% \bibliographystyle{apsrev4-1}
% \bibliography{cmprefs}


\end{document}